\documentclass{sf2a-conf}
\usepackage{graphicx}
\newcommand\ion[2]{{#1}~\textsc{{#2}}}
\newcommand\arcsec{\mbox{$^{\prime\prime}$}}%

\begin{document}

\TitreGlobal{SF2A 2006}

%%-----------------------------
%%      the top matter
%%-----------------------------
\title{Plasma diagnostic of a solar prominence from hydrogen and helium resonance lines}
\author{Labrosse, N.}\address{Institute of Mathematical and Physical Sciences, University of Wales Aberystwyth, UK}
\author{Vial, J.-C.}\address{Institut d'Astrophysique Spatiale, CNRS--Universit\'e Paris Sud, Orsay, FR}
\author{Gouttebroze, P.$^2$}
\runningtitle{Solar prominence diagnostics}
\setcounter{page}{237}
% Keep this line, even if the page will be settled afterwards..
\index{Labrosse, N.}
\index{Vial, J.-C.}
\index{Gouttebroze, P.}
% Repeat the authors here, this will help to make the final index

\maketitle
\begin{abstract}
  We present the first comparison of profiles of H et He resonance lines observed by SUMER with theoretical profiles computed with our non-LTE radiative transfer code. We use the \ion{H}{i} Lyman~$\beta$, \ion{H}{i} Lyman~$\epsilon$, and \ion{He}{i} $\lambda$\,584~\AA\ lines.  
  Our code allows us to obtain the plasma parameters in prominences in conjunction with a multi-line, multi-element set of observations.
The plasma temperature in the prominence core is $\sim 8600$~K and the pressure is 0.03~dyn cm$^{-2}$. The Ly$\beta$ line is formed in a higher temperature region (more than 11\,000~K). 
\end{abstract}
%
%%-----------------------------
%%      your text
%%-----------------------------
\section{Introduction}
  
  The H and He resonance lines have a strong diagnostic potential. However, observations are difficult to interpret as the prominence plasma is optically thick at these wavelengths and out of LTE. It is thus necessary to use a non-LTE radiative transfer code to analyse the data.
  
\section{Observations and modelling}

\begin{figure}[h]
   \centering
 \includegraphics[width=7cm]{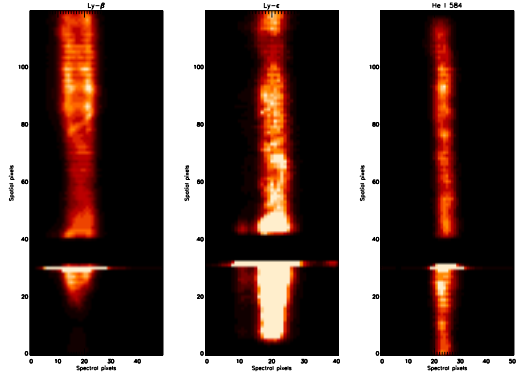}
  \caption{Intensity in the Ly$\beta$, Ly$\epsilon$, and \ion{He}{i} 584 lines as a function of wavelength and position along the slit.}
  \label{images}
\end{figure}

Observations were made during the 13th MEDOC campaign (IAS) on 2004, June 15. The SUMER slit centre was located at $X=870\arcsec$, $Y=465\arcsec$. 
%Ly$\beta$ was observed with the $0.3\arcsec \times 120\arcsec$ slit, and Ly$\epsilon$ and \ion{He}{i}~584 with the $1\arcsec \times 120\arcsec$ slit.
On Fig.~\ref{images} we note the det.~A anomaly around pixel 30: a dark area surrounded by bright pixels. We select one large area between pixels 70 and 100 to represent the prominence plasma. The Ly$\beta$ line exhibits a central reversal on several pixels. 
%The main features of the three observed lines are summarised in Tab.~\ref{compobs} and compared with the SUMER atlas observations of Parenti et al. (2005).

%\begin{table}[h]
%  \caption{Intensity and FWHM of the three lines compared with the SUMER atlas values. Units are $I$ (mW m$^{-2}$ sr$^{-1}$), FWHM (m\AA).}
%  \medskip
%  \label{compobs}
%  \centering
%  \begin{tabular}{ccccc}
%    \hline\hline
%    & \multicolumn{2}{c}{24/06/2004} & \multicolumn{2}{c}{SUMER atlas}\\
%    \cline{2-3} \cline{4-5}
%    {Line} & {I} & {FWHM} & {I} & {FWHM}\\
%    \hline
%    \ion{H}{i} Ly-$\beta$ & 235 & 474 & 244 & 396 \\
%    \ion{H}{i} Ly-$\epsilon$ & 8 & 323 & 13 & 312\\
%    \ion{He}{i} 584 & 154 & 127 & 152 & 96\\
%  \end{tabular}
%\end{table}

The numerical code solves the equations of radiative transfer and statistical equilibrium for H, \ion{He}{i}, and \ion{He}{ii}. Details are given in Labrosse \& Gouttebroze (2001, 2004). The prominence is represented by a 1D vertical plane-parallel slab. We consider 2 types of prominence atmospheres: constant temperature and pressure within the prominence, or with a prominence-to-corona transition region (PCTR). In the latter case the temperature and pressure profiles are of the same type as proposed by Anzer \& Heinzel (1999). 
%The model parameters (temperature, pressure, slab width, altitude, microturbulent velocity) are chosen in order to reproduce the variety of quiescent solar prominences.

\section{Results}

\begin{figure}[h]
   \centering
  \includegraphics[width=9cm]{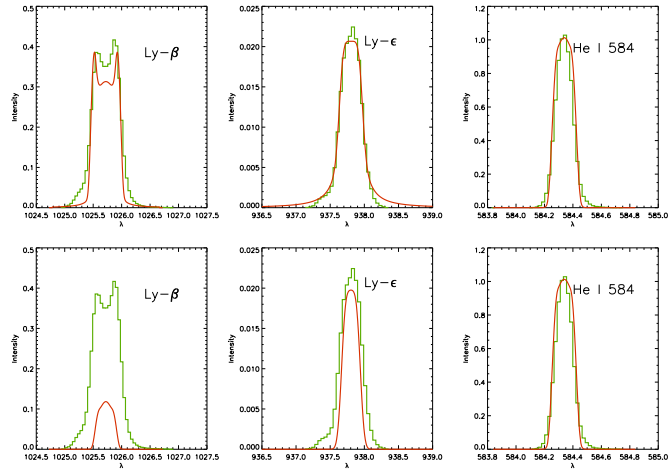}
  \caption{Comparisons between observed (green) and theoretical (red) profiles. Intensities are in W m$^{-2}$ sr$^{-1}$ \AA$^{-1}$.}
  \label{fit6}
\end{figure}

The top plots on Fig.~\ref{fit6} show each line profile observed by SUMER and the theoretical profile which best fits the observed profile. We obtain a different model for each line (see Tab.~\ref{results}). Each one of the three selected models is isothermal and isobaric. The PCTR models considered here cannot reproduce the observed profiles for this  prominence. 
We also searched for a unique model that simultaneously reproduces the three line profiles. The result is shown on the three bottom panels of Fig.~\ref{fit6}. The selected model is the same as the one which best fits the observed \ion{He}{i} 584 line alone (our model \# 3), but it fails to reproduce the Ly$\beta$ line. There is a long-standing problem regarding the Ly-$\beta$ line (see, e.g., Patsourakos \& Vial 2002). Some possible explanations are: \textbf{1)} We may need a different type of temperature profile for this particular prominence. \textbf{2)} The fine structure of the prominence should be incorporated in the modelling. \textbf{3)} The incident radiation illuminating the structure is strongly affected by the presence of several active regions on the disk seen on MDI images of the same day.

\begin{table}[h]
  \caption{Parameters for selected models. Parameters: temperature $T$ (K), pressure $p$ (dyn\,cm$^{-2}$), total column mass $M$ (g\,cm$^{-2}$), slab thickness $D$ (km), microturbulent velocity $V_t$ (km\,s$^{-1}$), height above the limb $H$ (km).}
  \medskip
  \label{results}
  \centering
  \begin{tabular}{cccccccccc}
    \hline\hline
    {Model} & {Line} & {$T$} & {$p$} & {$M$} & {$D$} & {$V_t$} & {$H$} & {$\tau_{912}$} & {$\tau_{504}$}\\
    \hline
    1 & Ly-$\beta$ & 11155 & 0.075 & 3.48~10$^{-6}$ & 584 & 19 & 3627 & 2 & 1\\
    2 & Ly-$\epsilon$ & 6677 & 0.414 & 1.12~10$^{-5}$ & 127 & 13 & 73198 & 26 & 3\\
    3 & 584 & 8602 & 0.029 & 4.50~10$^{-7}$ & 128 & 18 & 94333 & 0.5 & 0.1\\
    \hline
  \end{tabular}
\end{table}

\section{Conclusion}

The plasma temperature in the core of the observed prominence is found to be $\sim 8600$~K and the pressure is 0.03~dyn cm$^{-2}$. The Ly$\beta$ line is formed in a higher temperature region (more than 11\,000~K). The \ion{He}{i} line at 584 \AA\ has been succesfully used for the diagnostic of the prominence. The combination of the helium resonance lines with the Lyman lines of hydrogen provides new constraints for prominence models.

\end{document}